\pdfoutput=1

\documentclass[11pt]{article}

\usepackage[]{EMNLP2023}

\usepackage{times}
\usepackage{latexsym}
\usepackage{amsmath,graphicx}
\usepackage[T1]{fontenc}

\usepackage[utf8]{inputenc}

\usepackage{microtype}

\usepackage{inconsolata}
\usepackage{arydshln}
\usepackage{multirow}
\usepackage{placeins}
\usepackage{booktabs}

\makeatletter
\newcommand{\printfnsymbol}[1]{%
  \textsuperscript{\@fnsymbol{#1}}%
}
\makeatother

\usepackage{multicol}

\usepackage{bera}
\usepackage{listings}
\usepackage{xcolor}

\usepackage[normalem]{ulem}
\useunder{\uline}{\ul}{}

\definecolor{background}{HTML}{EEEEEE}

\lstdefinelanguage{json}{
    basicstyle=\normalfont\ttfamily,
    numbers=left,
    numberstyle=\scriptsize,
    stepnumber=1,
    numbersep=8pt,
    showstringspaces=false,
    breaklines=true,
    frame=lines,
    backgroundcolor=\color{background},
    literate=
     *{0}{{{\color{numb}0}}}{1}
      {1}{{{\color{numb}1}}}{1}
      {2}{{{\color{numb}2}}}{1}
      {3}{{{\color{numb}3}}}{1}
      {4}{{{\color{numb}4}}}{1}
      {5}{{{\color{numb}5}}}{1}
      {6}{{{\color{numb}6}}}{1}
      {7}{{{\color{numb}7}}}{1}
      {8}{{{\color{numb}8}}}{1}
      {9}{{{\color{numb}9}}}{1}
      {:}{{{\color{punct}{:}}}}{1}
      {,}{{{\color{punct}{,}}}}{1}
      {\{}{{{\color{delim}{\{}}}}{1}
      {\}}{{{\color{delim}{\}}}}}{1}
      {[}{{{\color{delim}{[}}}}{1}
      {]}{{{\color{delim}{]}}}}{1},
}

\title{E2E Spoken Entity Extraction for Virtual Agents}
 \author{Karan Singla \\ Whissle \\  karan@whissle.ai \AND Yeon-Jun Kim \\ Interactions-AI \\ ykim@interactions.com \And Srinivas Bangalore \\ Interactions-AI \\ sbangalore@interactions.com}

\begin{document}
\maketitle

\begin{abstract}
In human-computer conversations, extracting entities such as names, street addresses and email addresses from speech is a challenging task. In this paper, we  study the impact of fine-tuning pre-trained speech encoders on extracting spoken entities in human-readable form directly from speech without the need for text transcription. We illustrate that such a direct approach optimizes the encoder to transcribe only the entity relevant portions of speech ignoring the superfluous portions such as carrier phrases, or spell name entities. In the context of dialog from an enterprise virtual agent, we demonstrate that the 1-step approach outperforms the typical 2-step approach which first generates lexical transcriptions followed by text-based entity extraction for identifying spoken entities.

\end{abstract}

\section{Introduction}



Enterprise Virtual Agents (EVA) provide automated customer care services that rely on spoken language understanding (SLU) in a dialog context to extract a diverse range of intents and entities that are specific to that business \cite{price2020improved}. Gathering various entities like names, email, street address from human callers become a part of large range of virtual agents. In order to minimize the error in recognition and extraction of names, designers of speech interfaces often design prompts that request the user not only to say their name but spell it as well to address issues of homophones. (eg. {\em Catherine} or {\em Katheryn}). Such a behavior to spell carries over to other entities such as street and email addresses, without the users being explicitly prompted to do so.

\begin{figure}[thb]
\centering
\includegraphics[width=80mm,height=50mm]{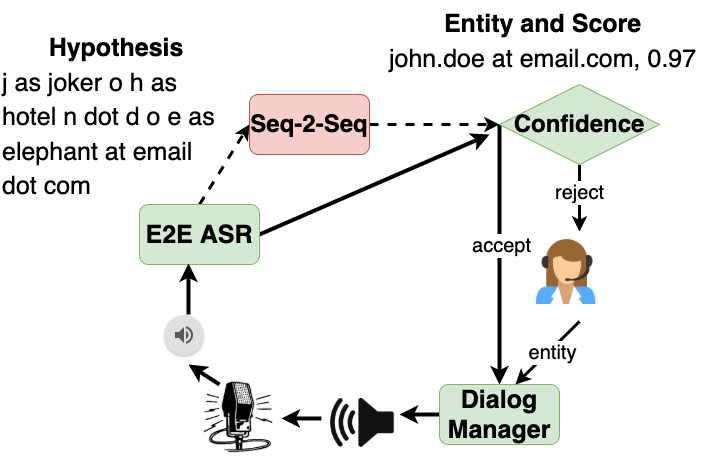}
\caption{Overview of our proposed EVA system with human-in-the-loop for entity extraction.}
\end{figure}

Extensive research has been done to recognize entities in spoken input \cite{favre2005robust,bechet2004detecting,sudoh2006incorporating,gupta2005att,kim2000rule}. Similar to text-based NER, approaches for Spoken NER often involve predicting entity offsets and type in text provided by an automatic speech recognizer (ASR) \cite{ghannay2018end, palmer2001improving, kubala1998named} or recognizing directly as a part of E2E ASR output. Significantly limited research has been done on spoken entity extraction in dialogs \cite{kurata2012leveraging, kaplan2020may}, and even fewer in enterprise virtual agents \cite{bechet2004detecting, gupta2005att}. Some methods proposed include using a predefined list of entity names~\cite{price2021hybrid} in a speech recognizer, fuzzy refinement by exploiting knowledge graphs~\cite{das2022listen}, or to using a large vocabulary speech recognizer to obtain the transcript further processed using text-based NER tools. Such techniques are difficult to adapt to caller responses to the prompt "say and spell your first/last name" in a spoken dialog system, as illustrated by the following example.

\begin{lstlisting}[language=json,firstnumber=1]
s as in sam k as in kipe i as in ina ia b as in boy o as in over --> skibo
\end{lstlisting}

Until recently, where \cite{singla2022seq} adapt a standard Seq-2-Seq architecture to extract person names from automatically transcribed text. However, this \emph{2-step} approach means two systems to maintain and adapt, but also, possibly loss of acoustic-prosodic information. In this paper, we propose a novel method for extracting human-readable spoken entities \emph{directly} from speech with a single model (\emph{1-step} approach) that is optimized for the entity extraction task. We hypothesize CTC loss \cite{graves2006connectionist}, widely used for training E2E ASR systems in an non-autoregressive manner, can be re-imagined to map audio events to text events. By generating only entity relevant tokens, our system learns to perform more intelligent entity extraction, instead of just performing literal lexical transcription as done by all existing ASR systems.  We believe this opens the door to more interesting use-cases where E2E ASR systems learns to perform task-specific generation directly from speech.

We acquire data from a production EVA system which has human-in-the-loop for automation and data-collection purposes. Section 3 describes dataset in detail. We found that our proposed  1-step approach significantly outperforms the 2-step approach for extracting names, address and emails from users of an EVA system.  The contributions of our paper are as follows:

\begin{itemize}
    \item We adapt standard E2E ASR architectures optimized using CTC loss and inferred using greedy decoding to transcribe only entity relevant tokens directly from speech.
    \item We show that our proposed method performs better when compared to 2-step cascading approach and also better than human annotators in a fully automated human-in-the-loop dialog system.

\end{itemize}

\section{Related Work}
A common practice is to convert normalized token sequence in spoken form produced by ASR into a \emph{written form} better suited to processing by downstream components in dialog systems \cite{pusateri2017mostly}. This written form is then used to extract \emph{structured information} in the form of intent and slot-values to continue a dialog \cite{radfar2020end}. Recently, there is a growing tread to use neural encoders optimizing directly using speech input, popularly known as \emph{E2E SLU} approaches \cite{serdyuk2018towards,haghani2018audio}.

{\bf Inverse Text normalization: }Information extraction systems generally use an Inverse text normalization (ITN) component to convert a token sequence in spoken form produced by ASR into a written form suitable for downstream components -- NLU and dialog. Transforming spoken language to written form involves altering entities like cardinals, ordinals, dates, times, and addresses \cite{sak2016written, pusateri2017mostly}. Methods proposed for ITN include: using language models (LM) to decode written-form hypothesis \cite{sak2016written},  a finite-state verbalization model \cite{sak2013language}, leveraging rules and handcrafted grammars to cast ITN as a labeling problem \cite{pusateri2017mostly}. 

{\bf E2E SLU:} Several E2E approaches which directly act on speech, have been proposed for named entity recognition, a closely related task to the entity extraction studied in this work \cite{ghannay2018end,tomashenko2019recent, caubriere2020named,yadav2020end,shon2022slue}. Ghannay et al.\cite{ghannay2018end} fine-tune an E2E ASR pre-trained with the CTC loss \cite{amodei2016deep} with a set of special character labels enclosing the named entities in the transcription using CTC.
Others works also re-purpose E2E encoders pretrained for ASR pretraining followed by NER fine-tuning with tagged sequences \cite{amodei2016deep, ghannay2018end, caubriere2020named, tomashenko2019recent, yadav2020end, shon2022slue}. 

Unlike these previous works that typically need both the text transcript along with entity type and offset tags, our approaches only need the normalized entity for supervision. Our system transparently only use pairs of audio and the target normalized entities to extract. Thus, removing a significant amount of cost and effort needed to obtain transcription and entity tags.



\section{Method}

We rethink ASR not only to be a transcription system but an E2E speech based encoder that can extract human-readable entities thus, learning to ignore, normalize and generate only target entity tokens directly from speech. 

\subsection{Non-Autoregressive Speech Based Extraction}

We re-purpose an E2E ASR fine-tuned for standard transcription task and fine-tune it using {\em (Speech-input, Entity)} pairs using CTC loss \cite{graves2006connectionist}. It does this by summing over the probability of possible alignments of input steps to target entity relevant tokens, producing a loss value which is differentiable with respect to each input node. We use NeMo library \cite{kuchaiev2019nemo} for all training and testing purposes.

In this work, we pick an off-the-shelf Citrinet \cite{majumdar2021citrinet} model downloaded from NeMo library \footnote{https://tinyurl.com/ykzmwhre}. It is trained on a 7k hour collection of publicly available transcribed data and uses SentencePiece \cite{kudo2018sentencepiece} tokenizer with vocabulary size of 1024 ($L$)\footnote{https://tinyurl.com/y3n9drj2}. We fine-tune it again for the transcription task using additional 800 hrs of transcribed speech from a collection of enterprise virtual agent applications. This model achieves a word accuracy of 93.1\% on a 28k utterance test set consisting of user utterances that are in response to ``How may I help you?'' opening prompt from various enterprise virtual agent applications \cite{singla2022seq}.

E2E Citrinet ASR is then re-purposed and fine-tuned for entity extraction. For entities (email and postal addresses) which contain vocab tokens not part of ASR tokenizer (digits, special symbols) we initiate the classification head using a sentence piece tokenizer learnt on fine-tuning data. Vocabulary size is kept as 1024 in all experiments. We fine-tune this E2E encoder for direct entity extraction from speech using a standard CTC loss. 

\subsubsection{From Network output to Entities}

 We use the same mathematical formulation as CTC \cite{graves2006connectionist} to classify unseen speech input sequences to minimise task specific error measure. Similar to standard practice, our CTC network has a softmax layer with one more unit than there are labels in $L$. The activation of the extra unit is the probability of observing a 'blank' or no label. The activation of the first L units are interpreted as the probabilities of observing the corresponding labels at particular times. But contrary to standard interpretation of CTC, our system output is contextualized over larger time-steps to output only entity relevant tokens. 

More formally, for an input sequence x of length T (steps in a sample) define a speech DNN encoder with m inputs, n outputs and weight vector w as a continous map $N_w: (R^m)^T \mapsto (R^n)^T$. Let $y = N_w(x)$ be the sequence of network output, and denote by $y_{k}^{t}$ the activation of output unit $k$ at time $t$. $y_{k}^{t}$ is interpreted as the probability of observing label k at time t, thus, defining a distribution over the set $L^{'^T}$ of length T sequences over the alphabet $L^{'} = L \cup \{blank\}$: 

\begin{equation}
    p(\pi|x) = \prod_{t=1}^{T} y_{\pi_t}^t 
\end{equation}

we refer to the elements of the $L^{'^T}$ as paths, and denote them $\pi$

\cite{graves2006connectionist} makes an implicit assumption in Equation 1 that outputs at different times are conditionally independent. However, feedback loops within encoders connecting different position information makes them conditionally dependent. One possibly important reason behind the success of CTC based E2E ASRs using convolutional or transformer blocks.

Many$-$to$-$one map $B$ is defined as $L^{'} \mapsto L^{\leq T}$, where $L^{\leq T}$ refers to set of sequences of length less than or equal to $T$ over the original label alphabet L. All blanks and repeated labels are removed from the paths. Thus when optimized to output only entity relevant tokens, system outputs blanks for those time-steps instead of mapping them to a token in $L$ (step-wise CTC outputs in Figure 2). Finally, $B$ is used to define the conditional probability of an entity $l \in L^{\leq T}$ as sum of probabilities of all the paths corresponding to it:

\begin{equation}
    \sum_{\pi \in B^-1(e)} p (\pi|x)
\end{equation}

\begin{figure*}
\centering

    \scalebox{0.18}{
    \includegraphics[]{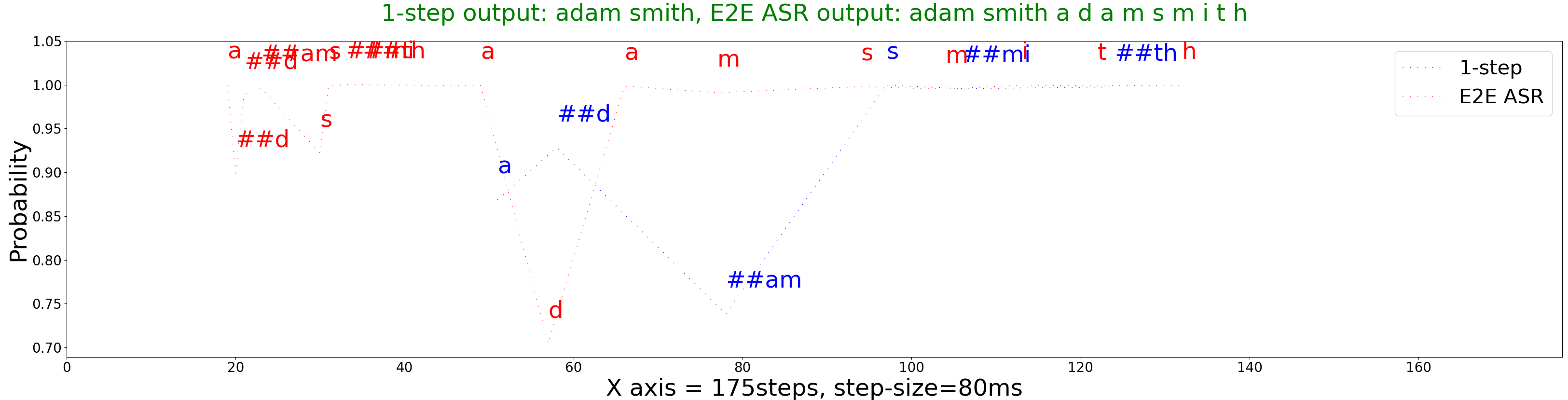}
    }

    \scalebox{0.18}{
    \includegraphics[]{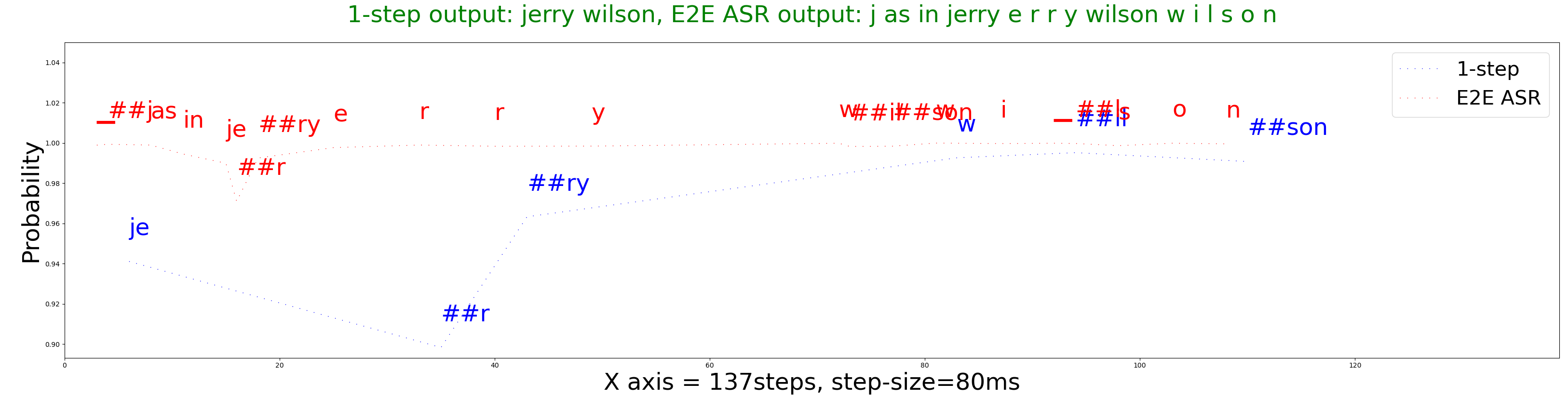}
    }

\caption{Samples showing output of greedy decode comparing an E2E Citrinet ASR with our proposed 1-step approach. Time steps which are not marked with any token are predicted as \emph{blanks}. Blue tokens which mark our 1-step system's output which learns to ignore and interpret non-relevant tokens.}
\end{figure*}

Figure 2 shows sample with output tokens at each time step (80ms for Citrinet) along with probability of that token. It shows system learns to ignore parts of speech to focus on spell, ignore phrases and also interpret \emph{j as in jeery -> j}. Thus, CTC loss helps to output contextualized tokens and also align them to steps in audio without any supervision. Our experiments suggest that other E2E ASR architectures, like Conformer \cite{gulati2020conformer} show similar results, when fine-tuned with non-autoregressive CTC loss.

At training time, classifier construction is done according to \cite{graves2006connectionist} and implemented in NeMo Library. We refer the reader to original Citrinet paper for more implementation details \cite{majumdar2021citrinet}. For decoding, we use simple greedy CTC decoding (best path method) where the $argmax$ function is applied to the output predictions and the most probable tokens are concatenated to form a preliminary output. CTC decoding rules i.e remove blank symbols and repeated tokens, are applied to obtain readable entity. 

\subsection{Baseline: Cascading ASR and NLU systems}

In this 2-step approach, we first transcribe the speech provided by humans into text using same pretrained E2E ASR checkpoint used by \cite{singla2022seq}. We then extract entities from the transcribed text by learning to translate using {\em (transcription, entity)} pairs. We use a standard off-the-shelf transformer based Seq-2-Seq system to extract entities from transcribed text baseline\footnote{https://github.com/mead-ml/mead-baseline}. We found using 4 multi-headed instead of 2 performs better for all entities. Table 1 shows sample input $H$ and desired output $M$.
 
The ASR hypothesis is provided in the form of byte-pair encoded (BPE) tokens as input to the Seq-2-Seq model, while the decoder generates entity relevant BPE tokens. We use a shared embedding layer for both encoder and decoder tokens. We use fastBPE\footnote{https://github.com/glample/fastBPE} to learn a shared vocabulary for both the encoder and decoder. We use Adam optimizer with a fixed batch size of 32 and a fixed learning rate of $1.0e-5$. We do not perform any pre-training of text-based seq-2-seq model but instead train our system from a random initialization. We provide confidence-score as the sum of log-probability assigned to the BPE entity sequence.


\section{Dataset}

While some public datasets like the OGI collection~\cite{cole1995new} include a small subset of spelled names. EVAs for multiple industry verticals, record millions of user utterances responding to different prompts. In these dialog systems customers are prompted to provide  information at various dialog turns using different authored prompts. Table~\ref{tab:auto} shows few sample responses (hypothetical imitation) from callers and desired system output.

\begin{table}[htb]
\centering
\scalebox{0.8}{
\begin{tabular}{|l|}
\hline
\multicolumn{1}{|c|}{{\bf What's your name?}} \\
\multicolumn{1}{|l|}{{\bf H: }a l e x u s last name k i n g}\\ 
\multicolumn{1}{|r|}{{\bf M: }alexus king}\\ 
\multicolumn{1}{|l|}{{\bf H: }l i s a s t a n t as in tom o n}\\
\multicolumn{1}{|r|}{{\bf M: }lisa staton}\\ \hdashline
\multicolumn{1}{|c|}{{\bf What's your street address?}} \\
{\bf H: }four three eight three remo crescent road\\
\multicolumn{1}{|r|}{{\bf M: }4383 remo crescent rd.}\\
{\bf H: }six forty six eighteenth street apartment one\\
\multicolumn{1}{|r|}{{\bf M: }646 state st. apt 1}\\
\hdashline
\multicolumn{1}{|c|}{{\bf What's your Email-id ?}} \\
{\bf H: }k as kite i n as nancy nine one five at gmail.com\\
\multicolumn{1}{|r|}{{\bf M: }kin915@gmail.com}\\

{\bf H: }a n as in nancy girl tower e e at outlook dot com\\
\multicolumn{1}{|r|}{{\bf M: }angtee@outlook.com}\\
 \hline
\end{tabular}
}
\caption{Sample responses from users of an EVA system when prompted with the question.}
\label{tab:auto}
\end{table}

We collect training data from several production EVA applications including banking, insurance, mobile service, and retail from callers based in the United States. Our collections are user responses in the form of audio samples and labels by human-in-the-loop agents. Human agents listen to customer inputs, then either type a human-readable entity or report an invalid input provided by a user. We remove the samples where user doesn't provide a meaningful inputs (keeping 70\%-85\% utterances). Thus creating a data with {\em (speech, entity)} pairs used by automatic extraction systems. Table \ref{tab:data} show statistics for each prompt type we use namely, first name, last name, full name, postal and email address for experiments. We keep additional valid sets which are size of 10\% of train sets for model selection. Table \ref{tab:data} also shows median duration in the training set and also 95\% percentile range for it.

\begin{table}[htb]
\centering
\scalebox{0.67}{
\begin{tabular}{|cccc|}
\hline
\textbf{Type} & \textbf{Dur (95\% perc.)} & \textbf{\#Train} & \textbf{\#Test}            \\ \hline

First name (FNAME)                     & 7.0s (3.8 - 15.6)                          & 89k                               & 835                   \\
Last name (LNAME)                   & 6.5s (4.0 - 13.3)                          & 522k                              & 1k  \\
Full name (FULLNAME)                     & 10.1s (3.4 - 21.1)                         & 241k                              & 1k \\
Street address (STREET)                & 6.5s (2.6 - 15.4)                          & 1.2m                              & 4.3k \\
Email address (EMAIL)                      & 12.8s (3.8 - 31.2)                         & 620k                              & 1k \\ \hline
\end{tabular}
}
\caption{Statistics of traning and evaluation set.}
\label{tab:data}

\end{table}

For testing purposes, we randomly sampled audio from a large pool of data, which is collected at different time frames than train data, but from same set of applications. We imitate an human-in-the-loop scenario where annotators listen to user inputs, type the entity by listening to audio only once in the limited time. Our test participants are a mix of native and non-native speakers who could be less exposed to {\it European} names.  Table \ref{tab:data} shows size for test data for each type. It is often observed that the test participants introduce errors when labeling in a constraint setting like an EVA. Later, we employ native speakers of English to verify and correct entity labels. The last column in Table \ref{tab:results} shows human-in-the-loop performance in a constraint setting. 

We merge transcriptions of training data obtained using an E2E ASR for all entities  to create a list of most frequent keywords (excluding characters, number words, email-address-providers). Table 3 shows 15 most frequent keywords and their total \% frequency normalized by total samples for each entity type. For example: word \emph{as} is used in providing Email $502.4$ times at average by a caller in 100 inputs. Callers take help of additional words and phrases most for providing Email address, followed by Fullname, and least for street address.

\begin{table}[ht]
\centering
\scalebox{0.62}{
\begin{tabular}{|cccccc|} \hline
\textbf{Keyword} & \textbf{FNAME} & \textbf{LNAME} & \textbf{FULLNAME} & \textbf{STREET} & \textbf{EMAIL} \\ \hline
as       & 9.8   & 27.8  & 60.6     & 0.4    & 502.4 \\
in       & 9.3   & 28.7  & 59.9     & 1.3    & 493.6 \\
apple    & 1.8   & 2.0   & 6.9      & 0.2    & 50.6  \\
nancy    & 1.1   & 1.7   & 4.3      & 0.0    & 25.9  \\
sam      & 0.6   & 1.9   & 2.9      & 0.0    & 21.3  \\
like     & 2.1   & 3.3   & 3.9      & 0.1    & 11.7  \\
tom      & 0.5   & 1.4   & 2.4      & 0.0    & 15.2  \\
elephant & 0.3   & 0.5   & 2.1      & 0.0    & 17.1  \\
mary     & 0.8   & 0.8   & 1.8      & 0.1    & 12.6  \\
boy      & 0.4   & 1.5   & 1.9      & 0.1    & 11.1  \\
dog      & 0.3   & 0.7   & 1.6      & 0.1    & 11.9  \\
edward   & 0.5   & 1.1   & 1.7      & 0.0    & 10.5  \\
igloo    & 0.0   & 0.2   & 1.4      & 0.0    & 12.0  \\
cat      & 0.1   & 0.5   & 1.2      & 0.0    & 11.2  \\
for      & 1.1   & 0.7   & 3.0      & 0.1    & 8.6   \\
\hline
\end{tabular}
}
\caption{Keywords (freq > 20k) and their total frequency for each entity normalized with total samples for that entitiy type in training set}
\end{table}

\section{Experiments and Evaluation}

We use only 1 NVIDIA A100 GPU for all fine-tuning purposes. We keep batch-size of 32 with starting learning rate of .001. We use weight decay of .001 and update the model with 8 accumulated batches with adam back-propagation algorithm. We will share our experiment configuration in the final revision. We report results for average of 5 runs. We provide entity confidence score by summing over the posterior probability of non-blank predicted tokens, a method originally proposed by \cite{kumar2020utterance}. 

\begin{table}[htb]
\centering
\scalebox{0.8}{
\begin{tabular}{|ccccc|}
\hline
 & \multicolumn{4}{c|}{\textbf{Accuracy (in \%)}}              \\
 Entity & {\ul 2-step} & {\ul 1-step} & {\ul 1-step-joint} & {\ul Human} \\
FNAME                                   & 85.0         & 86.9         &  {\bf 89.3}  & 84.0  \\
LNAME     & 89.0         & \multicolumn{1}{c}{{\bf 92.2}}         &   92.1  & 89.1 \\
FULLNAME & 65.6         & \multicolumn{1}{c}{ 77.1}         &  {\bf 82.4}   & 75.0\\
STREET & 77.8         & \multicolumn{1}{c}{{\bf 81.9}}         & 80.2 & 73.2    \\
EMAIL & 61.5         & \multicolumn{1}{c}{66.1}         & 68.2 & {\bf 73.6}     \\   \hline
\end{tabular}
}
\caption{Results for correctly extracting entities for test set}
\label{tab:results}

\end{table}


\begin{table*}
\centering
\scalebox{0.99}{
\begin{tabular}{|lll|}
\hline
\textbf{ASR Transcription} &
  \textbf{ASR → S2S} &
\textbf{E2E extraction} \\ 
jack smith j a k s m i t h           & \textcolor{red}{jack smith}      & \textcolor{green}{jak smith}         \\
fingh s i n g h                        & \textcolor{red}{fingh}      & \textcolor{green}{singh}         \\
lunscarard l u n d s t a a r d         & \textcolor{red}{lundstaard} & \textcolor{green}{lundsgaard}  \\
o leary o capital l apostrophe e a r y & \textcolor{red}{leary}      & \textcolor{green}{ol'eary}     \\
fourty one hundred twenty third street & \textcolor{red}{4100 23rd street} & \textcolor{green}{41 123rd street} \\ \hline
\end{tabular}
}
\caption{Few samples cases where our proposed 1-step approach performs better than 2-step approach. Text in \colorbox{red}{red} highlights output is wrong, while \colorbox{green}{green} is correct.}
\label{tab:emails}
\end{table*}

Results for our proposed 1-step extraction outperforms 2-step entity extraction approach, as shown at Table \ref{tab:results}. The difference in performance is significant (permutation tests, n = $10^5$, all p < 0.05). Results also show that our systems achieve better performance than human annotators for most prompt type except email addresses. We found extracting email addresses is hardest of all types. This is in line with our hypothesis Email-IDs are hardest to extract because of possibly infinite combinations humans can make to describe a unique ID (approximately 70\% of email training data used some form of carrier phrase like "as", "in" and "like"). Joint model which pools training data for all entities shows improved performance for first-name, full-name and email extraction.

\textbf{Varying amount of training data:} Our proposed approach depends upon supervised data for automation. We analyze the amount the data needed before the system starts showing results which are useful to replace humans in an EVA. Figure 2 shows variation in Accuracy for full-name extraction test set. We measure accuracy at the level of words i.e: word is either first name or the last name, and at the level of characters. We found that system achieves high accuracy at the level of characters with less training data but needs more data to get complete name correct. 

\begin{figure}[ht]
\centering
\scalebox{0.25}{
    \includegraphics[]{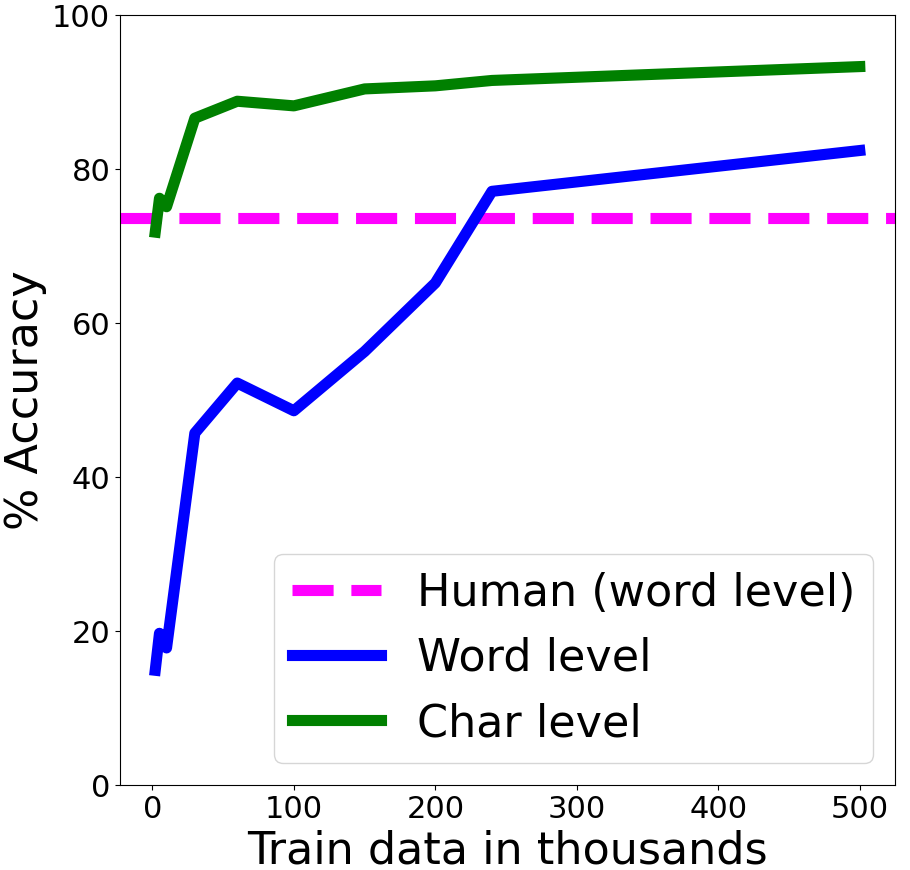}
}
    \label{fig:1}
\caption{Varying training data and measuring accuracy for 1-step approach.}
\end{figure}

\textbf{Effect of transcription quality:} Results in Table 5 show that performance of cascade approach is better when human transcribed text is fed to S2S system. Performance of 2-step S2S trained on the same noisy data as E2E extraction system seems more robust as it produces high quality results if correct transcriptions are provided. However, generating transcriptions with no errors is practically impossible and also acquiring data to fine-tune an ASR for this task will be costly.

\begin{table}[thb!]
\centering
\scalebox{0.99}{
\begin{tabular}{|c|ccc|}
\hline
\multirow{3}{*}{\textbf{Type}} & \multicolumn{2}{c}{\textbf{2-step}} & \textbf{1-step}    \\
                               & \multicolumn{2}{c}{Transcription}   & \multirow{2}{*}{-} \\
                               & {\ul Human}        & {\ul ASR}      &                    \\
First name                     & \textbf{89.0}      & 85.0           & 86.9               \\
Last name                      & \textbf{92.4}      & 89.0           & 92.0               \\
Steet address                  & \textbf{84.0}      & 77.8           & 81.9               \\ \hline
\end{tabular}
}
\caption{Comparing performance when human transcribed text is used instead of ASR output.}
\end{table}

\section{Observations}

{\bf Linguistic analysis:} We found humans break their answer into spell with or without language descriptions e.g: s as in sam more for email than other entities. Table 4 shows output for both 2-step and 1-step approach for extracting entities. We found cascading approach using S2S performs better if transcribed text provided by ASR has less errors. We believe some of these errors in transcription are due to pre-bias in language of ASR training data vocabulary. Improved performance of E2E extraction system indicates it can learn to resolve ambiguities for efficient entity extraction.
\\
{\bf Automation Rate:} Virtual agents use confidence score provided by an automatic module to decide whether call should be routed through a human agent. The confidence threshold determines the error versus rejection curve and a suitable operating point is chosen that optimizes the rejection at a given error rate. Figure 4 shows error rejection for fullname extraction. Our 1-step approach shows 12\% error rate at 20\% rejection, while 2-step approach shows 25\% error at 20\% rejection. It also performs better than human-in-the-loop at 20\% rejection rate.

\begin{figure}[bht]
\centering
\scalebox{0.25}{
    \includegraphics[]{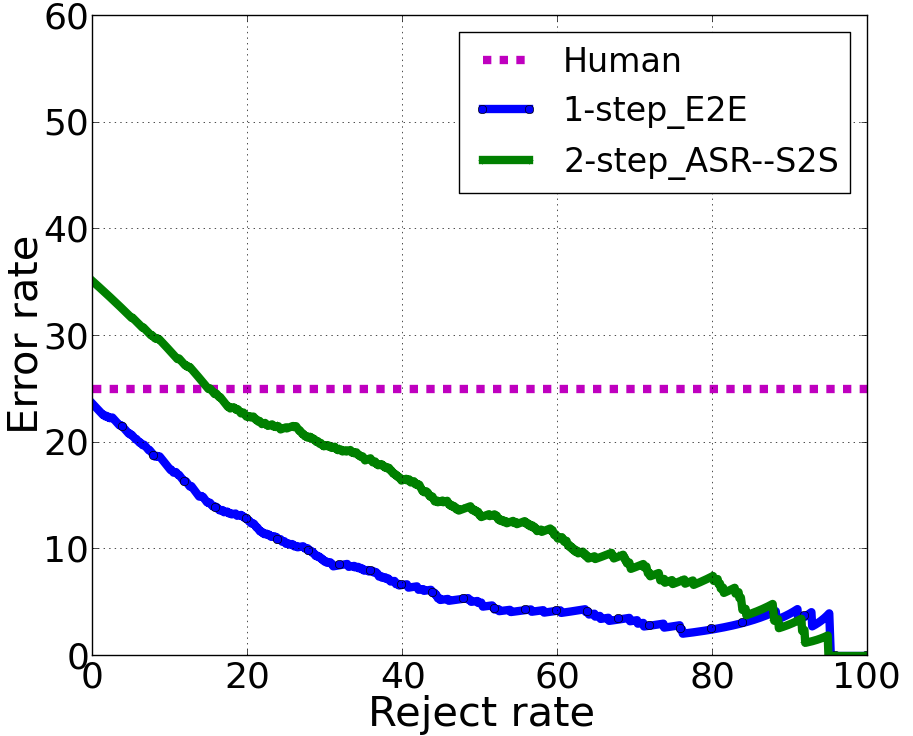}
}
\caption{Error-rejection curves for full name extraction. Setting a threshold helps dialog system designer control automation rate.}
\end{figure}

\section{Conclusions}
In this paper we show high-quality spoken entities can be extracted directly from speech by fine-tuning E2E ASR systems. The proposed 1-step model may not be influenced by ASR mistakes while carrying the critical token sequence to the final entity extraction phase. We didn't do hyper-parameter search for the models, due to GPU limitations.

For complete automation of prompts in customer calls a system also needs to extract intent for samples (10-15\%) with no entities in it. Our early experiments suggest this can be done by mixing intent labelled data (intent label used as single vocab token) or transcriptions of samples with no entities along with entity extraction data. This leads to minor loss in performance for each entity.




\bibliography{emnlp2023}
\bibliographystyle{acl_natbib}

\appendix

\end{document}